%% file: heterogeneous_replica.tex
\documentclass{sig-alternate-05-2015}

 \usepackage{subfigure}
  \usepackage{booktabs} % For formal column families
  \usepackage{amssymb}
  \usepackage{graphicx}
  \usepackage{amsmath}
  \newcommand{\subparagraph}{}
  \usepackage{titlesec}
  \usepackage{mathrsfs}
  \usepackage{makeidx}
  \usepackage{multirow}
  \usepackage{algorithm}
  \usepackage{pgfplots}
  \usepackage{framed}
  \usepackage{algorithmic}
  \graphicspath{ {images/} }
  \usepackage{url}
    \newtheorem{definition}{Definition}
  
  \usepackage{enumerate}

\begin{document}

% Copyright
\setcopyright{acmcopyright}

% DOI
\doi{10.475/123_4}

% ISBN
\isbn{123-4567-24-567/08/06}

%Conference
%\conferenceinfo{PLDI '13}{June 16--19, 2013, Seattle, WA, USA}

%\acmPrice{\$15.00}

%
% --- Author Metadata here ---
\conferenceinfo{WOODSTOCK}{'97 El Paso, Texas USA}

\title{Benchmarking Time Series Databases with IoTDB-Benchmark for IoT Scenarios}%

\numberofauthors{1} %  in this sample file, there are a *total*

\author{
% 1st. author
\alignauthor
Rui Liu, Jun Yuan, Xiangdong Huang\\
       \affaddr{Tsinghua University}\\
       \affaddr{Beijing, China}\\
       \email{245422695@qq.com, richard\_yuan16@163.com, huangxdong@tsinghua.edu.cn}
}

\maketitle
\begin{abstract}
With the wide application of time series databases (TSDB) in big data fields like cluster monitoring and industrial IoT, there have been developed a number of TSDBs for time series data management. Different TSDBs have test reports comparing themselves with other databases to show their advantages, but the comparisons are typically based on their own tools without using a common well-recognized test framework. To the best of our knowledge, there is no mature TSDB benchmark either. With the goal of establishing a standard of evaluating TSDB systems, we present the \emph{IoTDB-Benchmark} framework, specifically designed for TSDB and IoT application scenarios. We pay close attention to some special data ingestion scenarios and summarize 10 basic queries types. We use this benchmark to compare five TSDB systems: IoTDB, InfluxDB, OpenTSDB, KairosDB and TimescaleDB. Our benchmark framework/tool not only measures performance metrics but also takes system resource consumption into consideration.
% Another important aspect of our benchmark framework/tool is that it takes system resource consumption into consideration and supports test process monitoring and test data management. 

\end{abstract}



%
%  Use this command to print the description
%
\printccsdesc

% We no longer use \terms command


\section{Introduction}

With the pervasive application of time series databases (TSDB) in big data fields, such as industrial IoT \cite{strohbach2015towards}, manufacturing and power net, various time series databases have sprung up, including InfluxDB \cite{series2015metrics}, KairosDB \cite{hawkins2017kairos}, TimescaleDB \cite{vstefancova2018evaluation} and OpenTSDB \cite{sigoure2012opentsdb}. In the space of such a large variety of TSDBs, how to choose the most appropriate database service that suits the business needs becomes an important issue for developers and IT managers. Therefore, a flexible benchmark tool that can effectively assess and compare the performance of different TSDBs is desired.

Traditional database benchmark, like the TPC-benchmark family \cite{tpc}, provides a wide range of benchmarks customized for specific application scenarios 
% (e.g. TPC-DI \cite{poess2014tpc} for data integration) 
along with corresponding official benchmark tools. However, there is currently no widely recognized nor well-designed benchmark or tool on the market particularly for TSDB. Besides, most existing benchmark tools do not support the management of configuration parameters or system resource monitoring data during the tests, let along persist or help analyze test results.

TPCx-IoT \cite{tpc-iot} in TPC-benchmark family claims to be the first industry benchmark for time series oriented systems. It aims at comparing different software and hardware solutions for IoT gateways. However, the scenarios are not suitable for many practical use, including batch out-of-order data ingestion, aggregation or down-sampling querying. 

%Besides, the corresponding benchmark tool, which is impelementd based on YCSB\cite{ycsb}, limits its flexibility because YCSB was mainly designed for NoSQL systems.

%The scenarios are highly related to TSDB system 

%but lacking flexibility and not suitable for our practical needs including batch out-of-data ingestion. Besides, this benchmark has not been widely used because 
%Its official benchmark suit is based on modified YCSB\cite{ycsb}.

Although a few time series based benchmark tools have been developed recently, most of them lack the ability to simulate diverse workloads of time series oriented applications. For example, TSDBBench uses YCSB-TS \cite{ycsb-ts-github}, an extended version of YCSB \cite{cooper2010benchmarking}, as its official benchmark tool. Because YCSB is designed specifically for NoSQLs, the tool is hard to simulate flexible time series workloads, such as out-of-order data ingestion, time series with irregular frequency, and different data distributions. Other tools, such as InfluxDB-comparison \cite{influxdb-comparisons}, have similar drawbacks. All of these motivate us to establish a benchmark and tool to evaluate individual TSDB systems.
%The tool uses basic time domain functions, but is hard to simulate complex workloads.

%Further more, the workloads of YCSB consist of four fundamental operations, insert, query, update and delete, which only target on single record. While the tool supports five basic workloads based on different record selection distributions and operation ratios, these workloads all mix multiple types of operations, which introduces unpredictable impact on each basic operation because different kinds of operations can have interaction with each other within the same run (e.g., an update operation causes the delay of concurrent query operations due to storage locking). For the developers who care about the performance of a particular type of operation, this kind of workload makes it hard for analyzing individual basic operations and to locate the performance bottleneck.

%Although YCSB has provided an extensible interface to allow users to customize their own workloads, it still needs much effort to do so and it has limitations on workload parameters, for example, it is not able to generate out-of-order data workload that obeys a certain distribution. In addition, there are many new scenarios and needs that traditional benchmarks did not consider in TSDB application (e.g. time-series data distribution simulation). All of these motivate us to establish a standard benchmark to evaluate individual TSDB systems.

In this paper, we present IoTDB-Benchmark \\ (https://github.com/thulab/iotdb-benchmark) that is specifically designed for time series databases. First, the benchmark requires specifying various data distribution, because many TSDBs apply different data encoding algorithms, such as RLE \cite{pountain1987run} and Gorilla \cite{pelkonen2015gorilla}, which have significantly different effects on different data distributions. Another reason is that data distribution varies significantly in different applications and benchmark should cover these different data distribution types. Second, the benchmark requires specifying two kinds of operations for data ingestion: out-of time order data ingestion or data in the time order for ingestion in batch. The two operations are common in time series oriented applications while most existing benchmarks are not considered both. Third, the benchmark requires specifying not only the data ingestion operations and the query operations, but also what operation system metrics need to be collected.

We designed a IoTDB-Benchmark tool to support the above features. First of all, the benchmark tool has a data generator for simulating various data distributions, including but not limited to square wave, sine wave and sawtooth wave with controllable Gaussian noise. Second, the data generator can generate either out-of-order or in-order data while considering batch data ingestion. Third, the tool can be configured for not only individual basic operations, such as ingestion and query, but also mixture of operations to simulate complex real world workloads. Fourth, the system resource consumptions, for example, CPU usage and memory usage, are recorded for analysis. Last but not the least, the tool provides a general approach to manage benchmark result data, including persisting test configuration and test data and analyzing test results. Using all the data recorded with the tool, we demonstrate the performance comparison of four TSDB systems: InfluxDB, KairosDB, OpenTSDB, and TimescaleDB under various workloads.

The organization of this paper is as follow, Section 2 depicts the typical scenarios in TSDB application. Section 3 discusses the fundamental workloads of our benchmark. Section 4 discusses what and how to measure the performance of a TSDB system. Section 5 introduces more details of our benchmark tool. Experiments in Section 6 compare performance of multiple TSDB systems. Section 7 examines related work, and Section 8 is our conclusions.
\section{Scenarios}
\label{sec_scenario}
Different from many data center monitoring applications in IT companies, we mainly focus on industrial scenarios, which have more complex workloads. For example, a wind power company operates several wind-farms, each of which has some wind turbines and there are many sensors on each turbine for measuring hundreds of working metrics. Some sensors, such as temperature, generate periodical data with burrs. Some other sensors generate on-off value data or even data without obvious rules. Normally, the number of time series is large. For example, a wind power company in China operates 300 wind farms with total of 30 thousand wind turbines and each turbine collects at least 100 working metrics every 5 seconds or 7 seconds. Therefore, there are 3 million ($30,000 \times 100$) time series in total for all the turbines in this company. 

Unlike many DevOps applications collecting data every minute, the sensors data are collected and sent to the data center in large variety of frequencies. Some sensors send data with a high frequency, e.g., 1000Hz. Some sensors apply several frequencies and switch the frequency at different work situations. Other sensors have even irregular frequencies. 

In most cases, the data of sensors in a device are sent in batch (namely a packet) to the data center for reducing unnecessary transmission overhead. Then the server who receives packets may use one {\it batch insert} operation of database to write several packets to obtain a better throughput.

%In real world, data of different devices are sent to the data center separately with its own communication session. There needs different clients to simulate individual terminals and each client is responsible for a particular set of devices. 

Although the time series data is generated in the time order, the timestamps of time series data sent to TSDB are likely out-of-order due to the network latency and asynchronous transmission operations. This means that timestamps of data are not strictly incremented when the data arrives time series database service. 

%In addition, each time series data comes with its specific noise pattern. All above require a benchmark that can represent the situations.

%The frequency that sensors send data is generally high, for example 1000's data points per second, while data managers may only need to care about data on second level. This requires the database system to support aggregation queries. So the query workload should contain aggregation queries. 

Besides the ingestion workload to TSDB, another part of the workload is the queries. First, the most common case is that the user wants to query the data of a certain time range. Second, user wants to know the average or maximum value of the data. Third, the user may query the data within a certain value range. Besides, the data may be stored every millisecond but user only cares about data in seconds, which needs  sampling query to access data in different time granularity. Last but not least, the combination of all these requirements should also be considered.

To this end, we will examine the existing time series data test scenarios in further details and design a benchmark test configuration and test framework for industrial application scenarios.

\section{Benchmark Workloads}
In this section, we describe the general idea and principles that IoTDB-Benchmark should consider. %As update and delete operations in relational database are usually not very important to a TSDB system, we only focus on inserting and querying operations for the moment. 
For the scenarios described in Section 2, we present a flexible and scalable benchmark workload generator that can adjust the workload for a wide range of TSDB applications. We will explain the main parameters that users can configure for data ingestion and data query.

\subsection{Data Ingestion}
\label{sec_data_ingest}
IoTDB-Benchmark allows several parameters to configure different workloads for data ingestion. Users can define: (1) the schema of time series; (2) the data distribution; (3) the behavior of the batch operation; (4) the timestamp distribution (i.e., the frequency); and (5) whether the data is ordered by the timestamp.

{\bf The schema of time series}: we use {\it device group}, {\it device} and {\it sensor} to represent a time series. Users can define the number of device groups, the number of devices in each group and the number of sensors on each device. For each time series, users can define its value type, e.g., float, double, integer or string.

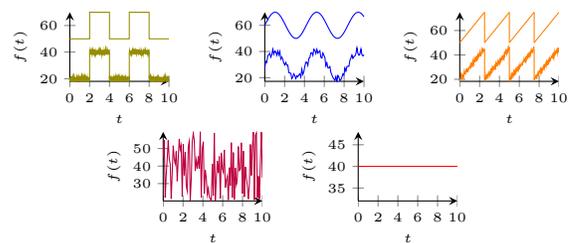
\begin{figure}[htbp]
\centering
\input{draws/series_square.tex}
\hskip 1pt
\input{draws/series_sine.tex}
\hskip 1pt
\input{draws/series_saw.tex}
\hskip 1pt
\input{draws/series_random.tex}
\hskip 1pt
\input{draws/series_constant.tex}
\caption{Five Types of Series Data Distribution}
\label{fig_5types_of_curve}
\end{figure}

{\bf The data distribution}: to better simulate the data in the real world, the benchmark defines the data generator to fit the periodic signal of the real scene according to some parameters.
% according to the set Fourier transform parameters. 
The generator supports five data distribution types for individual time series data: square wave, sine wave, sawtooth, random value within a certain range and constant value. In the first three types of data distribution, user can specify whether noise is added in the time series. Fig. \ref{fig_5types_of_curve} illustrates examples of the five types with and without noise.

\pgfplotsset{width=2.9cm, compat=1.9}

{\bf Batch operation}: as described in Section \ref{sec_scenario}, a TSDB usually receives data in batch. While each packet contains the data of all sensors in a device at certain timestamp, user can define how many packets are batch inserted in the target TSDB.

{\bf The timestamp distribution}: by default, each time series generates data in a fixed frequency. However, the benchmark also allows user to specify whether the frequency is irregular.

{\bf Out-of-order data}: there needs to have configuration parameters to control whether the timestamp is out of order and the proportion of out-of-order data. There are three different types of out-of-order data to consider: (1) Batch insert out-of-order: the data within each batch is out of order, but different batches are in order, that is, the timestamp of the next batch is always greater than the timestamp of the previous batch; (2) Global out-of-order: batches are not guaranteed to be ordered along time, that is, the overall data input is out-of-order; (3) Poisson-distribution out-of-order: the timestamp of the time series data is generated by following a certain probability distribution. The detail of this mode is introduced in Section \ref{sec_insert_test}.

\subsection{Query}
IoTDB-Benchmark supports ten query types. Suppose there is a relational table {\tt data (device, time, v1, v2, ..., v$n$)} ({\tt v$i$} is short for the value of sensor $i$) for storing time series data, then the 10 queries can be described as:
 \begin{itemize}
    \item Q1--Exact point query, i.e., {\tt select v1... from data where time=? and device in ?}.
    \item Q2--Time range query, i.e., {\tt select v1... from data where time $>$ ? and time $<$ ? and device in ?}.
    \item Q3--Query with limit and without filters, i.e., {\tt select v1... from data limit ?}.
    \item Q4--Time range query with value filter, i.e., {\tt select v1... from data where time $>$ ? and time $<$ ? \\ and v1 $op$ ? and device in ?}.
    \item Q5--Q4 with clause {\tt limit ?}.
    \item Q6--Aggregation query with time filter, i.e., {\tt select func(v1)... from data where device in ? and \\ time $>$ ? and time $<$ ?}.
    \item Q7--Aggregation query with value filter,  i.e., {\tt select func(v1)... from data where device in ? and \\ value $op$ ?}, where $op$ represents $>$, $<$ or $=$.
    \item Q8--Aggregation query with value filter and time filter, which is the combination of Q6 and Q7.  
    \item Q9--Latest point query, i.e., {\tt select time, v1...\\ where device = ? and time = max(time)}.
    \item Q10--Group by time range query. Group by time range is hard to be represented by a standard SQL, but is useful for time series data, e.g., achieving down sampling. Suppose there is a time series which covers the data in 1 day. By grouping the data by 1 hour, we can get a new time series which only contains 24 data points. There are more examples in Section \ref{query_test}. 
\end{itemize}
In the above queries, {\tt func()} represents an aggregation function, such as avg, min, etc., and {\tt ...} represents that there is more than one column in the {\tt select} clause.

%\subsection{Mix}
%The mixed workload consists of both ingestion workloads and query workloads. There are two ways to perform mixed load in benchmark tool.
%One is to use multiple benchmark test instances of different workloads performing simultaneously. Another way is to apply load generation module to generate a mixed load into the load pool, and the client threads take the load statement from the load pool to execute.

%In either way, the ratio of query load and write load can be configured with parameters.

\section{Performance Metric}
%A general relational database benchmark looks at two metrics: operation latency and space consumption, which are related to two important aspects in database design, performance optimization for different operations (read and write) and data storage space. The TSDB benchmark also needs to define what should be measured for TSDB and how to compare the results and rank performance across different TSDB systems.
\label{sec_metric}
% Like a general relational database benchmark measuring operation latency and space consumption for comparison across different RDBMS's, the TSDB benchmark also needs to define metrics for TSDB and how to compare the results and rank performance across different TSDB systems.
The performance of TSDB is evaluated by a set of metrics. First, a set of statistical metrics is needed to evaluate the performance of each type of operations, including minimum, maximum, average, middle-average, 1st, 5th, 50th, 90th, 95th and 99th percentile of \textbf{cost-time}. Cost-time is used as the performance measurement and it means the elapse time between sending a request or statement to the TSDB and receiving the full result from the TSDB successfully, which is also called latency or TTLB (Time to Last Byte). Middle-average is the average cost-time that cuts off 5\% head and tail. 

Second, we use \textbf{throughput} to evaluate the performance of ingestion test, which is calculated by the cost-time and the number of concurrent clients. We add up the ingestion cost-time of each client, respectively, as accumulative cost-time for each client and take the maximum accumulative cost-time as the total cost-time of multiple concurrent ingestion clients. The throughput equals to the total number of ingested data points divided by the total cost-time. 

%Third, the used disk space of the TSDB system during the test process (mainly in ingestion test) is monitored. We took the maximum difference between the start space consumption and used disk space during the test process as the \textbf{space consumption} of the TSDB. However, to accurately measure the disk space consumption is difficult because (1) most TSDBs use WAL \cite{wal}, which contains data under processing and it is difficult to predict when it will finish even if we can manually trigger the flush process and (2) some internal cache and in memory data are hard to measure too. Despite the inaccuracies, we use the disk consumption as the space consumption measurement for ingestion test.

Third, the used disk space of the TSDB system during the test process (mainly in ingestion test) is monitored. We took the maximum difference between the start space consumption and used disk space during the test process as the \textbf{space consumption} of the TSDB. That is, the space consumption may includes the file of Write Ahead Log (WAL) \cite{wal}, which is necessary for data recovery. Though the compression ratio of WAL files is smaller than the ratio of data files in many TSDBs, we consider the size of WAL is needed to be considered because it impacts how much total disk space an application needs.

%However, to accurately measure the disk space consumption is difficult because (1) most TSDBs use WAL \cite{wal}, which contains data under processing and it is difficult to predict when it will finish even if we can manually trigger the flush process and (2) some internal cache and in memory data are hard to measure too. Despite the inaccuracies, we use the disk consumption as the space consumption measurement for ingestion test.

Fourth, the \textbf{system resources consumption} during the test process is also considered. We measure several important system resource metrics including the system CPU, memory and disk I/O usage, memory used by TSDB service process, network receiving rate, disk I/O transfer number per second (tpc), disk write/read speed, etc.

\section{Benchmark Tool}
\label{sec_tool}
%We use a set of parameters to define a workload for TSDB system benchmark test. Although currently we focus on crucial operations as ingestion and query performance, we also measure other operation performance, like delete and schema registration, in automatic test process because they are needed to initialize and prepare the environment at some circumstances.

\subsection{The Process}
\label{sec_process}
\begin{figure}[htbp]
\centering
\includegraphics[height=2cm,width=8cm]{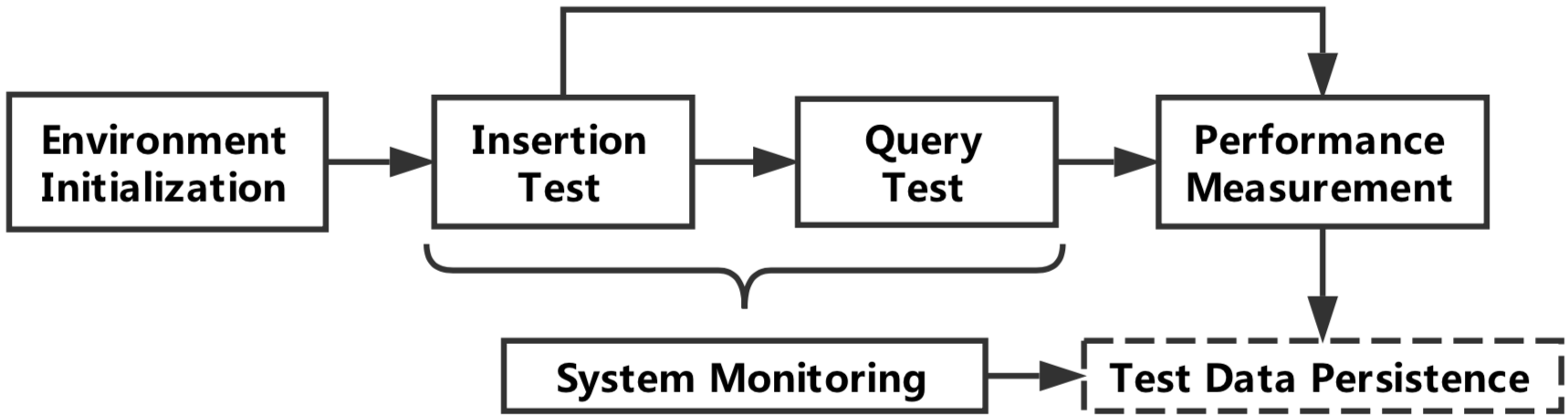}
\caption{Benchmark Test Procedures and Features Supported by the Benchmark Tool} 
\label{fig1}
\end{figure}

Fig.~\ref{fig1} shows the benchmark test process and other extensions that our benchmark tool supports. As shown in Fig.~\ref{fig1}, the whole test process can be divided into 6 phases, the box of solid line represents standard procedure of benchmark framework and the box in dotted line extended feature of the tool. 

\label{sec_arch}

\subsection{The architecture}
\label{sec_architecture}
\begin{figure*}[htbp]
\centering
\includegraphics[height=4.5cm,width=13cm]{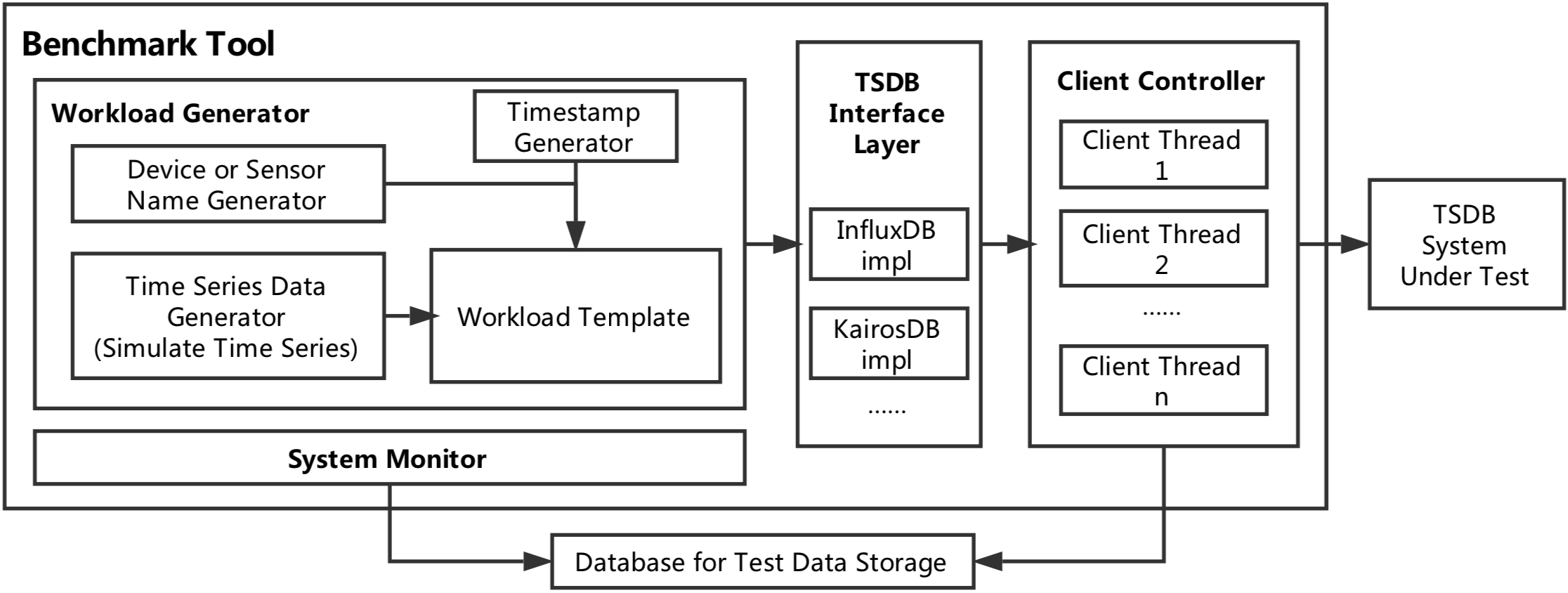}
\caption{Benchmark Tool Architecture} 
\label{fig2}
\end{figure*}

{\bf Environment initialization.} The first step of conducting a new test is to set up the test environment. In this step, users just need to configure parameters to define a workload, then our benchmark tool is able to automatically initialize the test environment, including starting the target TSDB service, removing old data and creating data schema if necessary. This initial setup is only executed before ingestion test. In another word, ingestion test starts with a cleaned, empty TSDB system. Besides, the old data cleanup process can be enabled or disabled through configuration.

{\bf Ingestion test.} After the environment setup, the ingestion test will be performed based on the configured workload. The details of the ingestion test are in Section~\ref{sec_insert_test}.

{\bf Query test.} Since the query test is based on the data schema created in the ingestion test, the query test is usually performed after ingestion test. The details of query test are in Section~\ref{query_test}.

{\bf Performance measurement.} The performance metrics such as cost-time are measured in each ingestion and query operation. When the entire test is completed, a set of statistical performance metrics introduced in Section 4 will be calculated. 
% All the performance measurement data will be logged as the test process goes.

{\bf System monitoring.} During the ingestion or query test process, the system resources will be measured as mentioned in Section \ref{sec_metric}. Through configuration, user can control the measuring frequency to fit different duration of test. 
%During the test, the system resources measurements data will be logged and output to an external database (e.g., a MySQL). 
Resources monitoring data can help us analysis different TSDB systems and provide another dimension of TSDB comparison.

{\bf Test data persistence.} Most benchmark tools prepare test results by simply output them to console or log them into files, which needs users' extra effort to analyze the data. The test data includes configurations used in each test instance, all the performance measurements, and the system resource monitoring data. Our benchmark tool allow users to use a relational database, such as MySQL, to store all the test data while the test goes. By using a database to manage the  test data, we can easily trace the test results, monitor test process and analyze the test data with SQL or better with a data visualization tool, like Tableau \cite{tableau} or Grafana \cite{grafana}. When performing long running test and there are many test instances running in parallel, this feature becomes absolutely necessary.

The above processes can be done automatically by our benchmark tool. Since there are many parameters can be configured, users can explore the performance impacts with these parameters. But the routine of manually configuring, initialization and launching test is tedious and exhausting. Our tool provide a simpler way to do a sequence of tests automatically by letting user edit a routine file, which defines what parameters should be altered before each test.

The modular design of the benchmark tool is shown in Fig.~\ref{fig2}. 
The idea of modularization makes our tool scalable and extensible to add new features and support new TSDB systems.  The tool contains 4 major parts: \textbf{workload generator}, \textbf{TSDB interface layer},  \textbf{client controller} and \textbf{system monitor}. 

{\bf Workload generator}: this module is responsible for generating ingestion or query SQLs or requests, which contains 4 sub-modules. 

When generating a new ingestion SQL or request, the {\it workload template} sub-module calls {\it timestamp generator} to get the next timestamp $t_i$ and fills the template with $t_i$ and the value generated by {\it series data generator}, which maintains a function of ($t_i$, $s_i$). Currently, 5 types of functions are supported (as listed in Fig. \ref{fig_5types_of_curve}, Section 3.1), and different sensors are assigned different value functions randomly from the above functions while  the appearance ratio of these function obeys user's definition. 
Besides, the parameters of the functions are user-defined, so sensors may have the same type of value function while the parameters such as the period,  the maximal value, and the offset varies from device to device. 

When generating a query SQL or request, the {\it workload generator}  generates time series list for a query, and then produces corresponding query clauses, such as the time/value filter and the aggregation function according to the query type. Then it fills the workload template with these values.

{\bf TSDB interface layer}: to support more TSDB systems, all the data ingestion and query operations are abstracted in system-free interfaces. By implementing these interfaces, users can apply the benchmark on more kinds of TSDB systems. 

%this module provides the templates of SQL statement or requests of different TSDBs and it is tightly connected with workload generator. The interface layer decides which implementation should be applied according to the targeting TSDB. It encapsulates different TSDB API designs and generates the templates with the same semantic, which matches our benchmark framework.

{\bf Client controller}: this module enables concurrent test through multiple client threads. The test client defines a higher level of test procedure, which consists of many basic operations that are implemented in the interface layer. This controller measures the cost-time of each operations and calculates the performance metrics when a test is complete. The performance metrics are stored into a relational database.% simultaneously with other test result data. %Besides, other errands like initialization and configuration storage are conducted by this module.

{\bf System monitor}: to fulfill the duties of monitoring, this module is developed for automatically measuring system resources previously listed and storing the data into the same relational database as that of test results during the test procedures. 

\subsection{Ingestion Test}
\label{sec_insert_test}
\subsubsection{Define Ingestion Workload.}

The ingestion workload can be configured by several parameters as shown in Table~\ref{tab1}. 
\begin{table*}
 
%\tiny
\scriptsize
%\footnotesize
\begin{center}
\caption{Main Ingestion Workload Parameters}\label{tab1}
\begin{tabular}{|l|l|l|}
\hline
Parameter name & Description & Example \\
\hline
GROUP\_NUMBER &  Total device group number, each group has several devices & 2\\
DEVICE\_NUMBER & Total device number & 10\\
SENSOR\_NUMBER & Sensor number per device & 3\\
CLIENT\_NUMBER & Concurrent client number & 5\\
BATCH\_SIZE &  Record number per batch & 100\\ % a record consists of a device's all sensor data with the same time-stamp. 
EPOCH &  Number of batch ingestion operations for each device  & 6\\
DATA\_TYPE & Data type of ingestion data & DOUBLE\\
POINT\_STEP & Milliseconds between two neighboring data points & 5000\\
TIMESTAMP\_GEN\_MODE & Decide how timestamp is generated & 0\\
IS\_MUL\_DEV\_BATCH & Decide if one batch contains data of different devices & False\\
IS\_RANDOM\_INTERVAL & Decide whether add a noise to POINT\_STEP & False\\
DISTRIBUTION\_RATIO & Ratio of five data distribution types & 1:1:1:1:1\\

\hline
\end{tabular}
\end{center}
 
\end{table*}

% Some companies only need to manage thousands of time series in the early stage and extend to millions of time series later. These parameters are set according to what an enterprise application needs and the test standard can flexibly support tests of different numbers of time series.

An example of ingestion workload parameters is listed in the rightmost column in Table~\ref{tab1} to illustrate the meaning of each parameter. Under the parameter set, the total number of time series to generate is 30 ($10 \times 3$), and the data from 10 devices are equally divided into 2 groups. The benchmark tool will use 5 client threads to send data ingestion requests to the TSDB service. Each client is bounded with a certain set of device evenly. It means client-thread-1 will only ingest the data of \emph{d\_0} to \emph{d\_1} and client-thread-2 \emph{d\_2} to \emph{d\_3}.

\noindent {\bf Data schema.}
%\subsubsection{Data schema.}
Every TSDB has its own data schema and we need to map the benchmark parameters to TSDBs' specific ones. In the case of InfluxDB, the concept of the device group corresponds to its {\it measurement}, and the device to its {\it tag}. We map the concept of the sensor to its {\it field}, which means there are three fields (\emph{s\_0}, \emph{s\_1} and \emph{s\_2}) in each measurement. Therefore, the tag \emph{device} take values from \emph{d\_0} to \emph{d\_4} in the measurement \emph{group\_0} and \emph{d\_5} to \emph{d\_9} in the measurement \emph{group\_1}. 
%For other TSDBs without the concept of the measurement or field, we design the data schema that suits the TSDB system best based on their official documents. 
For another example, we use {\it tag} to distinguish devices from sensors and regard a device group as a {\it metric} when it comes to OpenTSDB, because its official document shows that the concept of metric is a group of time series rather than a single time series. In fact, we did some experiments to compare the query performance of regarding group as metric or tag, and the result shows the former is better.

\noindent {\bf Ordered Data Ingestion.}
%\subsubsection{Ordered Data Ingestion.}
Still refer to the example in Table~\ref{tab1}. Each ingestion operation request (i.e. one batch) contains 100 records when BATCH\_SIZE=100. Each record has a single device's all sensors data with the same timestamp, just like a row in relational database, which has three data points in the example.

One batch in epoch $i$ for \emph{d\_j} is $B_{i_j}=[R_{0_{i_j}},R_{1_{i_j}},\cdots,R_{n_{i_j}}]$ where $n= {\rm BATCH\_SIZE} $. The subscript $i_j$ means batch $B_{i_j}$ belongs to epoch $i$ and device \emph{d\_j}, which means the data of one batch all belong to a single device. A record $R_{k_{i_j}}(timestamp,s_0,s_1,\cdots,s_m) = (t_{k_{i_j}},f_0(t_{k_{i_j}}),f_1(t_{k_{i_j}}),\\ \cdots,f_m(t_{k_{i_j}}))$ where $m = {\rm SENSOR\_NUMBER}$, $t_{k_{i_j}} = (i \times {\rm BATCH\_SIZE} + k) \times {\rm POINT\_STEP} $ and $f_m$ is the data distribution function assigned to $s_m$. When all devices complete one batch ingestion, it means epoch $i$ is done, and then the next epoch $i+1$ begins. For example, the first batch (i.e., epoch 0) for \emph{d\_j} is $B_{0_j}=[R_{0_{0_j}},R_{1_{0_j}},\cdots,R_{99_{0_j}}]$ where $R_{0_{0_j}}=(0,4.1,6.4,5.7), R_{1_{0_j}}=(5000,8.2,5.0,5.8),\cdots,R_{99_{0_j}} \\ = (495000,3.8,3.2,9.7)$.  Therefore, the total number of data points in one batch equals to SENSOR\_NUMBER $\times$ \\ BATCH\_SIZE. Since IS\_RANDOM\_INTERVAL is set to false and TIMESTAMP\_GEN\_MODE is 0 (meaning no out-of-order data), the timestamp of each sensor increases evenly. The whole ingestion test is completed when six epochs are done. 

When the insertion test is completed, each sensor/series has BATCH\_SIZE $\times$ EPOCH data points and the total number of data points is BATCH\_SIZE $\times$ EPOCH $\times$ SENSOR\_NUMBER $\times$ DEVICE\_NUMBER.

Write operations are often performed in batches in actual applications. We use parameter BATCH\_SIZE to specify batch-write operations. When BATCH\_SIZE = 1, the benchmark tool writes data point by point, which is equivalent to the write mode of many other test tools.

%\subsubsection{Out-of-order Data Ingestion}
\noindent {\bf Out-of-order Data Ingestion.}
The timestamp generating algorithm can be set by the TIMESTAMP\_GEN\_MODE parameter. There are three types of out-of-order modes, among which Poisson-distribution out-of-order mode is most commonly used. In this section we describe the specific implementation and mechanism to control the proportion $P$ of out-of-order data.

\begin{definition}
A time series ingestion workload $S$ is an ordered array of time-data tuple $T_i=(t_i,d_i)$ 
\begin{eqnarray}
\setlength{\abovedisplayskip}{-5pt}
\setlength{\belowdisplayskip}{-5pt}
S=[(t_0,d_0),(t_1,d_1),\cdots,(t_i,d_i),\cdots,(t_n,d_n)]
\end{eqnarray}
where the indicator $i$ is in the order the data arrive at TSDB, which means for any $i<j$, time-data tuple $T_i$ is sent to TSDB before $T_j$, or is previous to $T_j$ if in the same batch.

The out-of-order data is the time-data tuple $T_i$ which satisfy:
\begin{eqnarray}
t_i < \max \{\{t_j|0\leq j \leq i, j \in \mathbb{N}\}\}
\end{eqnarray}
\end{definition}
To calculate the timestamp of the next data point $t_i$, the benchmark tool maintains a maximum timestamp $CMT$ (Current Max Timestamp) of the currently written data. $t_i$ may be smaller than the current maximum timestamp $CMT$ by probability $P$, and how much smaller is decided by a random variable $X$, which obeys Poisson distribution. Otherwise, $t_i$ increases one step size, which happens by probability $1-P$. To formalize these descriptions, we give the following formula:
\begin{eqnarray}
\setlength{\abovedisplayskip}{-15pt}
\setlength{\belowdisplayskip}{-15pt}
t_i &=&
\left\{
\begin{array}{lll}
CMT - \Delta T, \ \ & P \\
CMT + {\rm POINT\_STEP}, \ \ & 1-P
\end{array}
\right.
\end{eqnarray}
where
\begin{eqnarray}
CMT &=& \max\{\{t_j|0\leq j \leq i, j \in \mathbb{N}\}\} \\
\Delta T &=& {\rm POINT\_STEP} \times (X+1) \\
X & \sim & Poisson(\lambda) \\
i.e.\ \  P(X=k) &=& \frac{\lambda^k}{k!} e^{-\lambda}, \ \  k \in \mathbb{N}
% i.e.\ \  P(X=k) &=& \frac{\lambda^k}{k!} e^{-\lambda}, \ \  k=0,1,2 \cdots
\end{eqnarray}

%Of course, the actual engineering implementation is much more complicated, including avoiding duplicate time-stamp and scale transformation in order to simulate real scenarios.

%The design of out-of-order data ingestion is based on the fact that there are often out-of-order data in the real world. For example, we have seen about 40\% of the data of a wind power company is out-of-order data, and the span of out-of-order time ($\Delta T$) is mainly distributed within 4 hours. Using Poisson distribution can appropriately simulate the out-of-order data time-stamp distribution of this scenario.

\begin{table*}
 
%\tiny
\scriptsize
%\footnotesize
\begin{center}
\caption{Main Query Workload Parameters}\label{tab2}
\begin{tabular}{|l|l|l|}
\hline
Parameter name &  Description & Example\\
\hline
QUERY\_TYPE &  Query workload type & 10\\
QUERY\_SENSOR\_NUM & Number of sensors involved in each query & 2\\
QUERY\_DEVICE\_NUM & Number of devices involved in each query & 2\\
QUERY\_AGG\_FUN &  Aggregation functions used by aggregation query &  max\\
CLIENT\_NUMBER &  Query client number & 2 \\
EPOCH &  Query requests number of each client & 100 \\
QUERY\_SPAN & Time filter span (ms)& 600000\\
QUERY\_VAL\_FILTER &  Query value filter & $>$ 0\\
TIME\_INTERVAL &  Interval of GROUP-BY-time query (ms) & 60000\\
\hline
\end{tabular}
\end{center}
 
\end{table*}

\subsection{Query Test}
\label{query_test}
\subsubsection{Define Query Workload}
Query load can be generated by randomly changing the parameter value in query filters for a particular type of query. By default, query test is based on the data generated by ingestion test. Therefore, the ingestion workload parameters related to data schema also affect query test. Apart from that, the main relevant parameters for query test are shown in Table \ref{tab2}:

Taking \emph{Q10-Group by time range query} as an example, the parameters are set as Table \ref{tab2} and the data set is generated by an ingestion test which uses configurations in Table~\ref{tab1}. The benchmark tool will use two client threads to send query requests to the target TSDB service.

Here, we use InfluxDB as an example to illustrate the actual query workload corresponding to these parameters. For other TSDBs, the query workload will be converted into equivalent operations.
The queries are of the following format:
\vspace{-10pt}
\begin{framed}
\vspace{-5pt}
\scriptsize{
\noindent
SELECT max({\bfseries \emph{s\_0}}), max({\bfseries \emph{s\_1}}) \\
FROM  {\bfseries\emph{group\_0}} , {\bfseries \emph{group\_1}} \\
WHERE ( device = '{\bfseries\emph{d\_3}}' OR device = '{\bfseries\emph{d\_8}}') \\ AND time $>=$ {\bfseries \emph{2010-01-01 12:00:00}} AND time $<=$ {\bfseries \emph{2010-01-01 12:10:00}} \\
GROUP BY time({\bfseries \emph{60000}}ms)
}
\vspace{-5pt}
\end{framed}
\vspace{-10pt}
\noindent where the sensors, devices and time range in the WHERE clause of each query are randomly selected.

This query means that each result point is the maximum of data in 1 minute span interval, which evenly divides the time range. Since the time range in query is 10 minutes, the query should return 10 aggregation result points for each data series. In this case, there are four series queried ({\bfseries \emph{s\_0}} and {\bfseries \emph{s\_1}} of {\bfseries \emph{d\_3}} in {\bfseries \emph{group\_0}} plus {\bfseries \emph{s\_0}} and {\bfseries \emph{s\_1}} of {\bfseries \emph{d\_8}} in {\bfseries \emph{group\_1}}), then the query should return 40 result points in total.

\subsection{Experimental Setup}
To reduce system complexity and random noise effect, we performed experiments on two server-class machines (two Intel Xeon CPU E5-2697 v4 @ 2.30GHz processors, 256-GB of memory, 10 disk RAID-5 array and 10 gigabit ethernet, the Operating System is Ubuntu 16.04.2 LTS 64-bit). One machine is used to run the TSDB service and the other the benchmark tool. Therefore, all tests are based on single-node TSDB installation. We monitor the system resources from client machine and give the client enough system resource so that the client machine wouldn't be a bottleneck.

In this paper, we benchmark four TSDB systems: InfluxDB 1.5.1, OpenTSDB 2.3.1 based on Hbase 1.2.8, KairosDB 1.2.1 based on Cassandra 3.11.3 and TimescaleDB 1.0.0 based on PostgreSQL 10.5. The configuration of each TSDB is tuned as much as we know to release its potential. We allocate enough and equal RAM or JVM space to each TSDB so that memory space wouldn't be a limitation. In particular, we set cache-max-memory-size and max-series-per-database of InfluxDB to unlimited. For OpenTSDB, we configured some parameters like tsd.http.request.enable\_chunked, \\ tsd.http.request.max\_chunk and tsd.storage.fix\_duplicates to enable large batch and out-of-order data ingestion. For KairosDB \cite{hawkins2017kairos}, the parameter read\_repair\_chance of Cassandra is set to 0.1 as official document instructed. Besides,  we use PGTune \cite{pgtune} to calculate the general optimized configuration of PostgreSQL for TimescaleDB.

% The data set sizes vary in different ingestion tests but query test runs on a uniform data set, which consists of 10000 time-series and each series contains 10000 data points with evenly increased time-stamp.

\subsection{Data Ingestion}

As introduced in Section~\ref{sec_data_ingest} and Section~\ref{sec_insert_test}, IoTDB-Benchmark provides many user configurable parameters to simulate real industrial time series data ingestion workloads. Our benchmark framework/tool enables users to explore the performance of TSDB systems and help them understand the relationship between performance and parameters, which is significant for system tuning. For ingestion test, we show several experiments that cover the five user-defined aspects mentioned in Section~\ref{sec_insert_test}, to show the tip of the iceberg.

\pgfplotsset{width=5.2cm, compat=1.9}
\begin{figure*}[htbp]
 
\centering
\subfigure[Client Number]{
\label{client_ingestion_results}
\input{draws/ingestion/concurrent.tex}
}
%\quad
\subfigure[Time Series Number]{
\label{series_ingestion_results}
\input{draws/ingestion/series.tex}
}
%\quad
\subfigure[Batch Size]{
\label{batch_ingestion_results}
\input{draws/ingestion/batch.tex}
}
%\quad
\subfigure[Data Distribution]{
\label{space_ingestion_results}
\input{draws/ingestion/space.tex}
}
%\quad
\subfigure[Timestamp Distribution]{
\label{step_ingestion_results}
\input{draws/ingestion/step.tex}
}
%\quad
\subfigure[Out-of-order Data Ratio]{
\label{overflow_ingestion_results}
\input{draws/ingestion/overflow.tex}
}
 
\caption{Ingestion Experiments. I, T, K, O are Short Names for InfluxDB, TimescaleDB, KairosDB, OpenTSDB Respectively (Throughput unit: 1000 points/second)}
\label{ingestion_results}
 
\end{figure*}
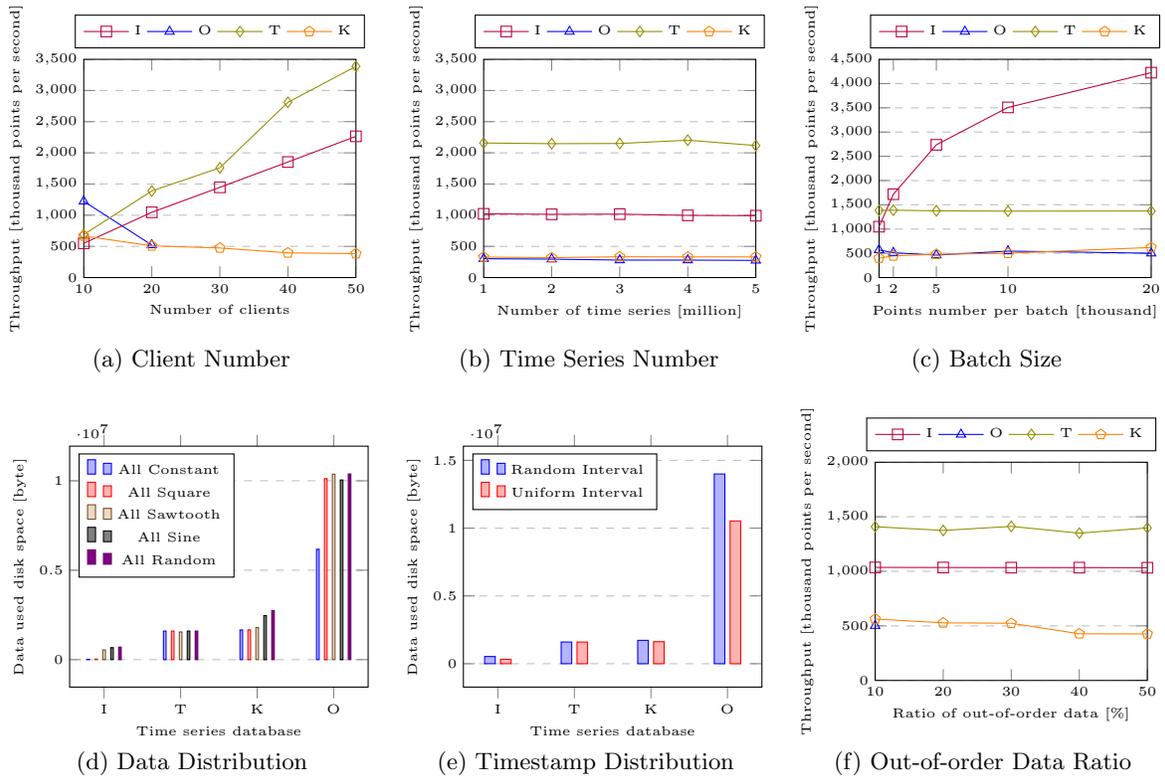

% {\bf Ordered Ingestion:}
% In most real industrial scenarios, time series data ingestion workload is in time order, which means the timestamp of data point the TSDB received are strictly incremental one after another. We performed five groups of experiments based on time-ordered insertion with changing of the following parameters: (1) the number of concurrent client; (2) the number of time series; (3) the batch size; (4) the data distribution; (5) the timestamp distribution.

{\bf Concurrent client number:} we compared the ingestion performance with different client numbers by configuring CLIENT\_NUMBER as shown in Fig.\ref{client_ingestion_results}. Except the client number, all TSDBs use the same configurations: 1000 devices in total and each device has 10 sensors of DOUBLE type. Each sensor/time-series will have 10000 data points (with BATCH\_SIZE=100, EPOCH=100) to be ingested and DISTRIBUTION\_RATIO=1:1:1:1:1. The configuration is regarded as a standard for other ingestion experiments.

The result shows that, with the client number increasing, the throughput of TimescaleDB and InfluxDB are growing while OpenTSDB and KairosDB are not. Especially for OpenTSDB, when the client number is bigger than 30, its ingestion performance drops dramatically so that the test can't even finish in the endurable time. Therefore, we only present the results of OpenTSDB when client number is fewer than 30.

{\bf The number of time series:} the time series number equals to DEVICE\_NUMBER $\times$ SENSOR\_NUMBER, therefore we set SENSOR\_NUMBER=100 and change DEVICE\_NUMBER to compare the ingestion performance under different time series number as shown in Fig.\ref{series_ingestion_results}. In this experiment group, we use 20 clients and enlarge the time series number to millions. We set BATCH\_SIZE=10 to make the data points number per batch the same with previous experiments in Fig.\ref{client_ingestion_results} since SENSOR\_NUMBER=100 and keep other parameters the same. 

Comparing Fig.\ref{series_ingestion_results} and Fig.\ref{client_ingestion_results} we can see that with the time series number increasing, the throughput of InfluxDB is almost the same.
% as in Fig.\ref{client_ingestion_results} when client number is 20, which means InfluxcDB has very good stability when scaling data schema.
TimescaleDB has even better performance when the sensors number in each record increases from 10 to 100, but the throughput of KairosDB and OpenTSDB dropped. From Fig.\ref{series_ingestion_results}, we can see that enlarging the series number has little impact on ingestion performance.

{\bf Batch size:} in Fig.~\ref{batch_ingestion_results} we compare the throughput with different BATCH\_SIZE. We use 20 clients and other parameters are the same as those in Fig.~\ref{client_ingestion_results}.

The result shows the throughput of InfluxDB grows quickly when the batch size increasing, which has limitation of 7 million points per second (not shown). While the throughput of other TSDBs have insignificant changes.

{\bf The data distribution:} as mentioned in Section \ref{sec_metric} our performance metric also involves disk space consumption, which is especially related with time series data distribution type. We set different DATA\_RATIO (e.g. 1:0:0:0:0 for all constant) to compare the performance when ingesting different types of time series data. Other parameters are the same. 

Results in Fig.~\ref{space_ingestion_results} show the disk usage efficiency of the four TSDBs, where InfluxDB $>$ TimescaleDB $\geq$ KairosDB $>$ OpenTSDB. For different distribution types, the disk space consumptions are also different, but the difference of throughput is insignificant (not shown).

{\bf The timestamp distribution:} to set IS\_RANDOM\\ \_INTERVAL to True can make the time interval of neighboring points irregular. Other parameters follow previous configurations. We compare the disk space consumption of the two scenarios as shown in Fig.~\ref{step_ingestion_results} and the result shows that the disk space consumption with uniform timestamp interval data is less than that of random timestamp interval data. Moreover, random time interval has little effect on ingestion throughput (not shown).

{\bf Out-of-order Ingestion:} the data of above ingestion tests are all in order of timestamp and we then alter to out-of-order data ingestion workload. We set TIMESTAMP\_GEN\_ \\ MODE=3, which is the Poisson-distribution out-of-order and change the ratio of out-of-order data (a parameter that not shown in Table \ref{tab1}) as shown in Fig.~\ref{overflow_ingestion_results}. Other parameters follow previous configurations.

The result shows that for InfluxDB, TimescaleDB and KairosDB, the throughput may slightly slow down when the out-of-order data ratio increases. While for OpenTSDB the out-of-order data leads to sharp atrophy of throughput, therefore we only present result of 10\% out-of-order data.

\pgfplotsset{width=5.2cm, compat=1.9}
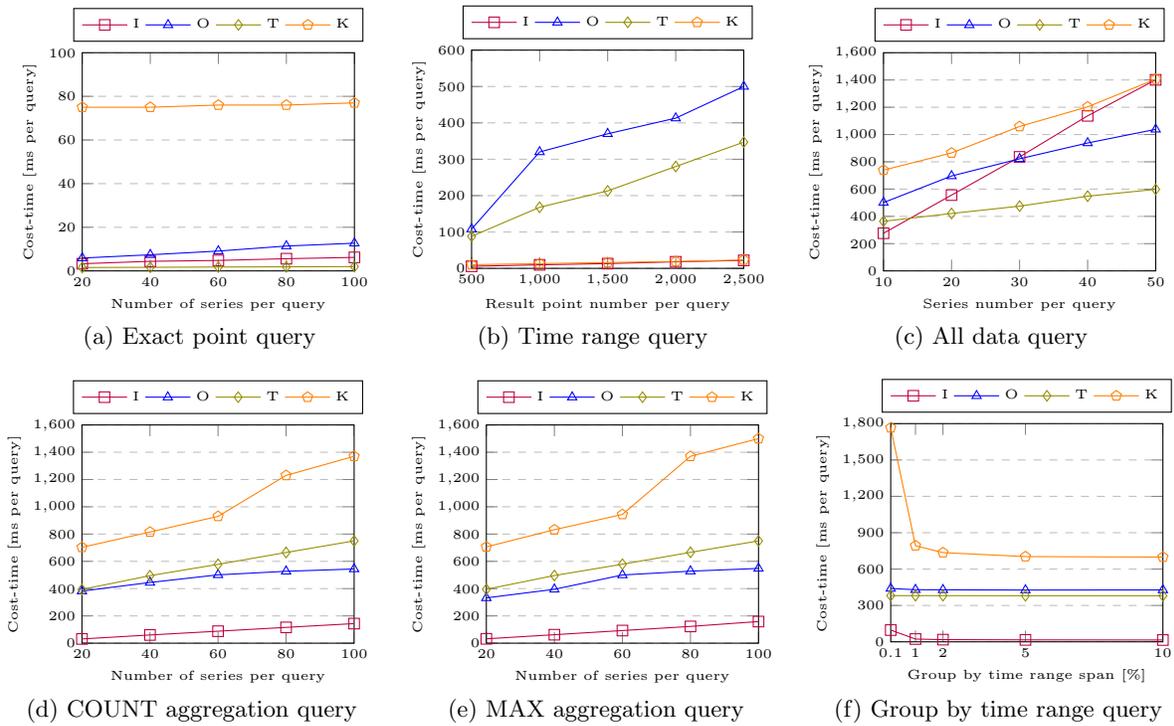
\begin{figure*}[htbp]
 
\centering
\subfigure[Exact point query]{
\label{exact_query_results}
\input{draws/precise_pic.tex}
}
%\quad
\subfigure[Time range query]{
\label{range_query_results}
\input{draws/range_pic.tex}
}
%\quad
\subfigure[All data query]{
\label{all_query_results}
\input{draws/collumn_pic.tex}
}
%\quad
\subfigure[COUNT aggregation query]{
\label{count_query_results}
\input{draws/count_agg_pic.tex}
}
%\quad
\subfigure[MAX aggregation query]{
\input{draws/max_agg_pic.tex}
%\caption{fig1}
}
%\quad
\subfigure[Group by time range query]{
\label{group_query_results}
\input{draws/groupby_pic.tex}
}
\setlength{\belowdisplayskip}{-3pt}
 
\caption{Four TSDBs's Average Cost-time of Different Query Workload Types}
\label{query_results}
% \vspace{-10pt}
\end{figure*}

\subsection{Query}

As introduced in Section 3.2, our benchmark supports 10 types of query. Because it is tediously long to illustrate all of them, we choose the typical four types of queries that cover the four basic query types in real scenarios listed in Section~\ref{sec_scenario}, as the query workload for benchmarking the same four TSDBs.

The four types are: Q1-Exact point query, Q2-Time range query, Q6-Aggregation query with time filter and Q10-Group by time range query.  Fig.~\ref{query_results} shows the results of the four TSDBs's average cost-time per query under different query workloads, in which Q1 is corresponding to sub-figure (a), Q2 is (b) and (c), Q6 is (d) and (e), and Q10 is (f). 
%Each point in the figure stands for a query test result of a query workload. 
Every query workload consists of 100 query requests and only one client.

% \titlespacing*{\subsection} {0pt}{3pt}{0pt}

{\bf Data Set:}
all these query tests are based on the data generated by a ingestion test, which contains 10 device groups. Each device group has 100 devices and each device has 10 sensors, therefore 10000 ($10 \times 100 \times 10$) time-series in total. Each sensor (i.e., time-series) contains 10000 data points with uniform time interval of 5 seconds.
%The time started at 2018/8/30T00:00:00.

\titlespacing*{\subsection} {0pt}{6pt}{0pt}
\subsection{System Resource Monitoring}
\begin{figure*}[htbp]

\centering
\includegraphics[height=8.3cm,width=12cm]{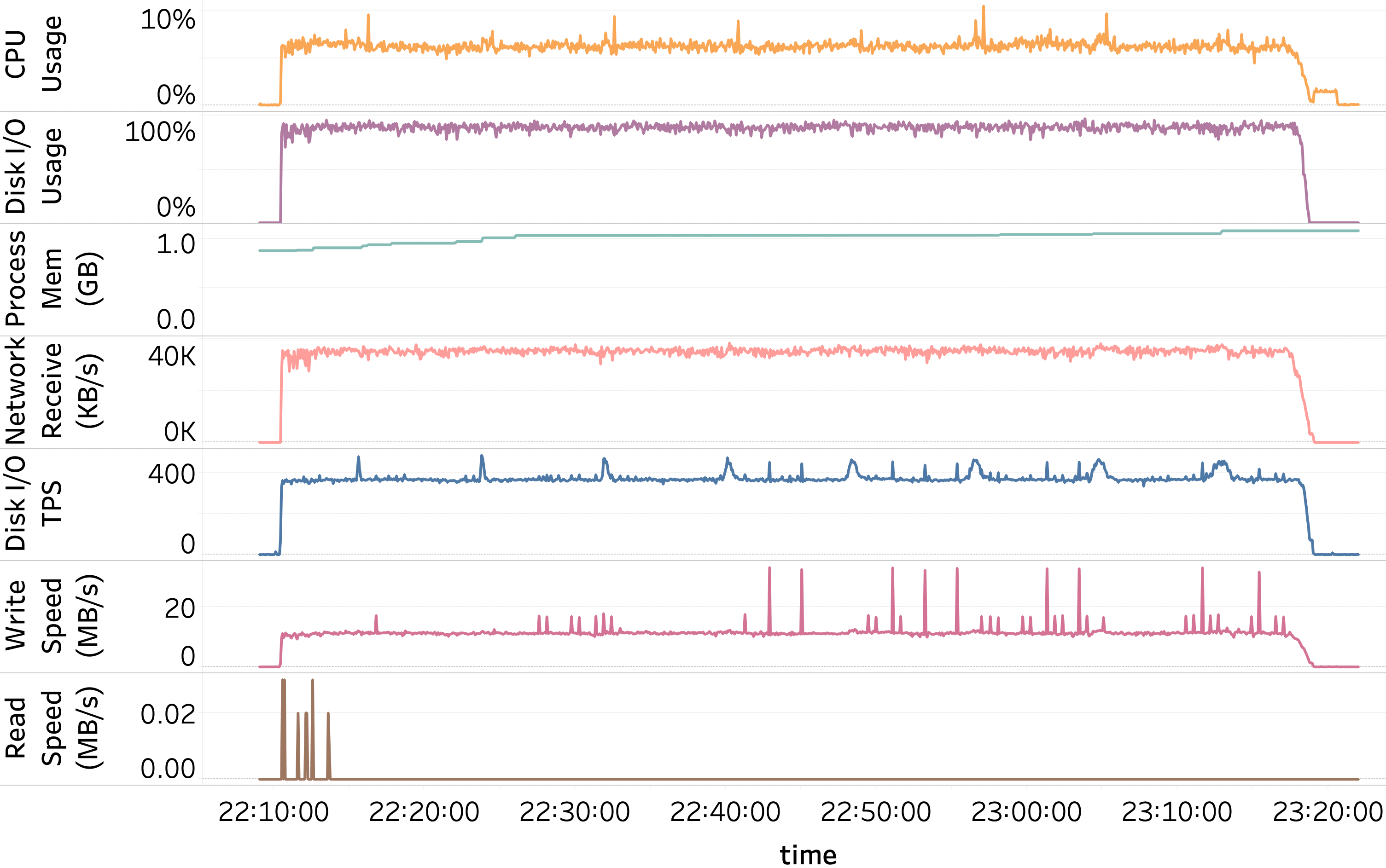}
\caption{System Resource Consumption Monitoring During InfluxDB Ingestion Test} 
\label{monitor}
 
\end{figure*}

{\bf Q1-Exact Point Query:}
when we compare the exact point query performance of series number changes in each query, we fix QUERY\_SENSOR\_NUM=10 and set QUERY\_ \\ DEVICE\_NUM = 2,4,6,8,10 respectively. We use the routine way introduced in Section 5.1 to execute this sequence of tests automatically and collect the results in Fig.~\ref{exact_query_results}. The result shows that the TimescaleDB is the fastest, and the next is InfluxDB and OpenTSDB, which are close to TimescaleDB, while KairosDB is the worst and about 35 times slower than TimescaleDB. This is because TimescaleDB is based on a relational database, PostgreSQL and uses time as the primary key, which is specially indexed. Besides, the effect of series number in each query is not significant.

{\bf Q2-Time Range Query:}
in Fig.~\ref{range_query_results}, we compare the time range query performance. When the time range changes in each query, it means the number of  each query's result points changes. We can see that the performance of InfluxDB and KairosDB are the best and the change of the time range has little influence on them. For TimescaleDB and OpenTSDB, the cost-time increased significantly with number result points increasing. In Fig.~\ref{all_query_results}, we set the time range to $50000000 ms$, which covers all timestamps, and change the series number in each query. We discover that the performance of InfluxDB is sensitive to series number comparing with other TSDBs.

{\bf Q6-Aggregation Query:}
we used COUNT and MAX as aggregation functions applied in Aggregation Query experiments. Each function is applied on different series number in each query as shown in Fig.~\ref{count_query_results} and (e). The results show that these two aggregation functions have almost the same performance for every TSDB. Besides, InfluxDB is the fastest, followed by OpenTSDB, TimescaleDB and KairosDB.

{\bf Q10-Group By Time Range Query:}
we use MAX as the aggregation function in the group by time range query and set QUERY\_SPAN=50000000, which covers the whole series. By configuring TIME\_INTERVAL, we compared the performance of different time range span as shown in Fig.~\ref{group_query_results}. For example, the {\it group by time range span} is 1\% means TIME\_INTERVAL is $1\% \times 50000000 ms = 500000 ms$. Therefore, each query will get $\frac{1}{1\%}=100$ aggregated points in the query result. We can see that the bigger the aggregation granularity, the better the performance, because the bigger aggregation granularity results in fewer result points. 

From the above query benchmark results, we can see that InfluxDB is leading the TSDBs since its performance in most query workloads are far better than the others. The query performance of OpenTSDB and TimescaleDB are close under many circumstances, while KairosDB is relatively weak, compared with other TSDBs, except the time range query. %或者这一段放到conclusion里

\noindent
Our benchmark  tool can not only collect test results, but also monitor the system resource consumption during the test process. We visualize the monitoring data listed in Section \ref{sec_metric} as shown in Fig.~\ref{monitor}, which monitors InfluxDB in Fig.~\ref{series_ingestion_results} when the series number is 1 million. We can see that the disk usage keeps on a high level while the CPU usage is about 6\%. 
It takes about one hour to run the whole test.
Due to limited space, the system resource consumptions of other TSDBs are not demonstrated.

% %
% \section{Future Work}

% 不太理解为什么有这么一章？？？
\section{Related Work}
{\bf Benchmark and Tool.}
% \subsubsection{Benchmark and Tool.}
% In the field of system testing, benchmark is a technology or standard that is widely used to evaluate the performance of computer hardware or software systems. Our work is both related to establishing TSDB benchmark and developing a benchmark tool to support it.
% 我们为什么又在下面这段里总结一遍？？？
% TPC is a classical benchmark family for narrow domain benchmarks like E-commerce of TPC-W. So far, TPC has released 19 standards including TPC-C, TPC-H and TPCx-IoT, which are related database. The TPCx-IoT is a benchmark for IoT gateways, which is similar to ours, but its official test suite is rather based on YCSB. Currently it only supports HBase and Couchbase-server. The TPCx-IoT workload includes data write and concurrent queries that simulate typical IoT gateway system workloads. Although TPCx-IoT is the first industry IoT standard, it still has some shortcomings. Since the official test suite is based on YCSB, it inherits some common problems of YCSB. For example, the workload configuration is not very flexible, and the batch write test and out-of-order data write test are not supported. 
Besides TPC benchmark and YCSB, there are other benchmarks and corresponding tools in the big data arena for our reference. YCSB-TS \cite{ycsb-ts-github} is a branch of YCSB that measures the performance of a TSDB. Andreas Bader \cite{bader2016comparison} used YCSB-TS to compare different TSDB systems. Since YCSB-TS is based on YCSB, it inherits some of the shortcomings of YCSB that have been encountered before. Ghazal \cite{ghazal2013bigbench}, et al, developed Bigbench to benchmark DBMS and its design ideas and extensibility are of research value. Chowdhury \cite{chowdhury2013bigbench} extends Bigbench to test Hadoop. Hibench \cite{huang2010hibench} is a benchmark tool developed by Intel for Hadoop. With the launch of Spark, BigDataBench, a big data benchmark tool, developed by UC Berkeley AMPLab Lab, was used to test big data systems, such as Spark. There are also other benchmark/tools that come with or associated to the database. For example \emph{influxdb-comparisons} is a comparison test tool written in Go language used by InfluxDB to compare InfluxDB with other NoSQL systems. However these benchmarks serve for specified products, which may lack justification for the compared systems.
% \\
% \\
%\subsubsection{Time series Database.}
\noindent  {\bf Time series Database.}
% While our benchmark too is targeting a time series database, the Time Series Database does not currently have a precise definition. A database management system (DBMS) with the 5 characteristics is generally considered to be a time series database\cite{tsdb-wiki}. Firstly, it stores records consist of timestamps, values and optional tags. Secondly, it stores multiple records grouped together. Thirdly, the data mode can be scaled flexibly. Besides, the query is usually based on time filter. Last but not least, the query can contain aggregation or down-sampling function.
The TSDBs we compared in this paper are considered as typical TSDBs because they covered 3 categories \cite{bader2017survey} of TSDB: (1) TSDB with no requirement on any DBMS; (2) TSDB with a requirement on NoSQL DBMS; (3) TSDB based on or modified from a relational database. InfluxDB, written in Go language, is one of the most popular time series data manage solutions. It has its own storage engine with TSM-Tree \cite{naqvi2017time}, which is an optimization of LSM-Tree \cite{o1996log}. OpenTSDB is a distributed and scalable TSDB based on HBase \cite{hbase2013apache}. Similar to InfluxDB, it uses tag to mark different series. KairosDB is forked from OpenTSDB, but is mostly based on Cassandra \cite{cassandra2014apache} storage while it also supports in memory storage called H2. Since Cassandra's rows are wider than HBase, KairosDB's Cassandra has a default row size of 3 weeks, while OpenTSDB's HBase is 1 hour. TimescaleDB is a TSDB based on relational database PostgreSQL  \cite{vohra2016using}, which still uses data schema of traditional relational database, but it is especially optimized for time-series data. It supports PostgreSQL's full SQL, which makes TimescaleDB inherit advantages of relational databases while meeting the needs of time-series data management.

\section{Conclusion}
In this paper, we present IoTDB-Benchmark for evaluating time series databases. The benchmark considers industrial IoT scenarios as the typical application of time series database and provides various parameters to simulate different scenarios and corresponding workloads in the real world. Correspondingly, we develop a benchmark tool to interpret those parameters, to conduct the benchmark testing, and to collect both test result data and system resource consumption data. We apply IoTDB-Benchmark and conduct several groups of experiments, including the ingestion test and query test on four popular TSDBs, InfluxDB, OpenTSDB, KairosDB and TimescaleDB. The results show the insights on them for developers and IT managers. The contributions and novel features of our work are as follow, first, we present a TSDB benchmark, which is especially for TSDB and designed for various application scenarios. Second, our benchmark takes system resource consumption metrics into consideration for recording, which is crucial for analyzing TSDB systems. Third, we apply relational database to manage the test data like performance metrics and corresponding configurations, which enable users to trace test results conveniently and conduct further analysis. With these features, IoTDB-Benchmark is able to provide benchmark for both development and research purpose. We will expand it for more TSDBs and add more workloads variations targeting on more complex scenarios in the future.

\bibliographystyle{abbrv}
\bibliography{heterogeneous_replica}

\end{document}

%% file: draws/series_square.tex
\begin{tikzpicture}
\tiny
\begin{axis}[
    axis lines = left,
    xlabel = $t$,
    ylabel = {$f(t)$},
]
%square
\addplot [
    domain=0:2, 
    samples=100, 
    color=olive,
    ]
    {20+2*rand};
\addplot [
    domain=2:4, 
    samples=100, 
    color=olive,
    ]
    {40+2*rand};
\addplot [
    domain=4:6, 
    samples=100, 
    color=olive,
    ]
    {20+2*rand};
\addplot [
    domain=6:8, 
    samples=100, 
    color=olive,
    ]
    {40+2*rand};
\addplot [
    domain=8:10, 
    samples=100, 
    color=olive,
    ]
    {20+2*rand};
\addplot [
    %const plot mark right,
    domain=0:10, 
    samples=100, 
    color=olive,
    ] coordinates {
        (0,50) (2,50) (2,70)
        (4,70) (4,50) (6,50)
        (6,50) (6,70) (8,70)
        (8,50) (10,50) 
    };
 
\addplot [
    %const plot mark right,
    domain=0:10, 
    samples=100, 
    color=olive,
    ] coordinates {
        (2,20) (2,40)
    };  
\addplot [
    %const plot mark right,
    domain=0:10, 
    samples=100, 
    color=olive,
    ] coordinates {
        (4,20) (4,40) 
    }; 
\addplot [
    %const plot mark right,
    domain=0:10, 
    samples=100, 
    color=olive,
    ] coordinates {
        (6,20) (6,40) 
    }; 
\addplot [
    %const plot mark right,
    domain=0:10, 
    samples=100, 
    color=olive,
    ] coordinates {
        (8,20) (8,40) 
    }; 
    
\end{axis}
\end{tikzpicture}

%% file: draws/series_sine.tex
\begin{tikzpicture}
\tiny
\begin{axis}[
    axis lines = left,
    xlabel = $t$,
    ylabel = {$f(t)$},
]
%Here the blue parabloa is defined
\addplot [
    domain=0:10, 
    samples=100, 
    color=blue,
    ]
    {10*sin(deg(x*1.5)) + 60};
%\addlegendentry{s\_1: sine}
\addplot [
    domain=0:10, 
    samples=100, 
    color=blue,
    ]
    {10*sin(deg(x*1.5)) + 30 + 3*rand};
\end{axis}
\end{tikzpicture}

%% file: draws/series_saw.tex
\begin{tikzpicture}
\tiny
\begin{axis}[
    axis lines = left,
    xlabel = $t$,
    ylabel = {$f(t)$},
]
%saw-tooth
\addplot [
    domain=0:2.5, 
    samples=100, 
    color=orange,
    ]
    {10*x + 50};
\addplot [
    domain=2.5:5, 
    samples=100, 
    color=orange,
    ]
    {10*x + 25};
\addplot [
    domain=5:7.5, 
    samples=100, 
    color=orange,
    ]
    {10*x};
\addplot [
    domain=7.5:10, 
    samples=100, 
    color=orange,
    ]
    {10*x - 25};
%\addlegendentry{s\_4: sawtooth}
\addplot [
    domain=0:2.5, 
    samples=100, 
    color=orange,
    ]
    {10*x + 20 + 2*rand};
\addplot [
    domain=2.5:5, 
    samples=100, 
    color=orange,
    ]
    {10*x - 5 + 2*rand};
\addplot [
    domain=5:7.5, 
    samples=100, 
    color=orange,
    ]
    {10*x - 30 + 2*rand};
\addplot [
    domain=7.5:10, 
    samples=100, 
    color=orange,
    ]
    {10*x - 55 + 2*rand};
%\addlegendentry{s\_4: sawtooth} 
\addplot [
    domain=0:10, 
    samples=100, 
    color=orange,
    ] coordinates {
        (2.5,20) (2.5,45)
    };  
\addplot [
    domain=0:10, 
    samples=100, 
    color=orange,
    ] coordinates {
        (5,20) (5,45) 
    }; 
\addplot [
    domain=0:10, 
    samples=100, 
    color=orange,
    ] coordinates {
        (7.5,20) (7.5,45) 
    }; 

\addplot [
    domain=0:10, 
    samples=100, 
    color=orange,
    ] coordinates {
        (2.5,50) (2.5,75)
    };  
\addplot [
    domain=0:10, 
    samples=100, 
    color=orange,
    ] coordinates {
        (5,50) (5,75) 
    }; 
\addplot [
    domain=0:10, 
    samples=100, 
    color=orange,
    ] coordinates {
        (7.5,50) (7.5,75) 
    }; 
\end{axis}
\end{tikzpicture}

%% file: draws/series_random.tex
\begin{tikzpicture}
\tiny
\begin{axis}[
    axis lines = left,
    xlabel = $t$,
    ylabel = {$f(t)$},
]

\addplot [
    domain=0:10, 
    samples=100, 
    color=purple,
    ]
    {rand*20 + 40};
%\addlegendentry{s\_1: sine}
\end{axis}
\end{tikzpicture}

%% file: draws/series_constant.tex
\begin{tikzpicture}
\tiny
\begin{axis}[
    axis lines = left,
    xlabel = $t$,
    ylabel = {$f(t)$},
    % ymin = 30,
    % ymax = 50
]

%Here the blue parabloa is defined
\addplot [
    domain=0:10, 
    samples=100, 
    color=red,
    ]
    {40};
%\addlegendentry{s\_1: sine}
\end{axis}
\end{tikzpicture}

%% file: draws/ingestion/concurrent.tex
\begin{tikzpicture}
\pgfplotsset{
    every axis legend/.append style={
    at={(0.5,1.05)},
    anchor=south
    }, 
}
\begin{axis}[
    legend columns=4,
    xlabel={Number of clients},
    ylabel={Throughput [thousand points per second]},
    xmin=10, xmax=50,
    ymin=0, ymax=3500,
    xtick={10,20,30,40,50},
    ytick={0,500,1000,1500,2000,2500,3000,3500},
    % legend pos=north west,
    ymajorgrids=true,
    grid style=dashed,
]
\addplot[
    color=purple,
    mark=square,
    ]
    coordinates {
    (10,549)(20,1046)(30,1447)(40,1853)(50,2265)
    };
\addplot[
    color=blue,
    mark=triangle,
    ]
    coordinates {
    (10,1222)(20,522)
    };
\addplot[
    color=olive,
    mark=diamond,
    ]
    coordinates {
    (10,690)(20,1388)(30,1760)(40,2810)(50,3387)
    };  
\addplot[
    color=orange,
    mark=pentagon,
    ]
    coordinates {
    (10,667)(20,508)(30,474)(40,396)(50,383)
    };
    \tiny
    %{{\legend{InfluxDB,OpenTSDB,TimescaleDB,KairosDB}}}
    {{\legend{I,O,T,K}}}
\end{axis}
\end{tikzpicture}

%% file: draws/ingestion/series.tex
\begin{tikzpicture}
\pgfplotsset{
    every axis legend/.append style={
    at={(0.5,1.05)},
    anchor=south
    }, 
}
\begin{axis}[
    legend columns=4,
    xlabel={Number of time series [million]},
    ylabel={Throughput [thousand points per second]},
    xmin=1, xmax=5,
    ymin=0, ymax=3500,
    xtick={1,2,3,4,5},
    ytick={0,500,1000,1500,2000,2500,3000,3500},
    %legend pos=north west,
    ymajorgrids=true,
    grid style=dashed,
]
\addplot[
    color=purple,
    mark=square,
    ]
    coordinates {
    (1,1022)(2,1014)(3,1017)(4,997)(5,992)
    };
\addplot[
    color=blue,
    mark=triangle,
    ]
    coordinates {
    (1,302)(2,295)(3,280)(4,280)(5,275)
    };
\addplot[
    color=olive,
    mark=diamond,
    ]
    coordinates {
    (1,2158)(2,2147)(3,2150)(4,2203)(5,2117)
    };  
\addplot[
    color=orange,
    mark=pentagon,
    ]
    coordinates {
    (1,328)(2,319)(3,333)(4,331)(5,331)
    };
    \tiny
    %{{\legend{InfluxDB,OpenTSDB,TimescaleDB,KairosDB}}}
    {{\legend{I,O,T,K}}}
\end{axis}
\end{tikzpicture}

%% file: draws/ingestion/batch.tex
\begin{tikzpicture}
\pgfplotsset{
    every axis legend/.append style={
    at={(0.5,1.05)},
    anchor=south
    }, 
}
\begin{axis}[
    legend columns=4,
    xlabel={Points number per batch [thousand]},
    ylabel={Throughput [thousand points per second]},
    xmin=1, xmax=20,
    ymin=0, ymax=4500,
    xtick={1,2,5,10,20},
    ytick={0,500,1000,1500,2000,2500,3000,3500,4000,4500},
    %legend pos=north west,
    ymajorgrids=true,
    grid style=dashed,
]
\addplot[
    color=purple,
    mark=square,
    ]
    coordinates {
    (1,1049)(2,1718)(5,2737)(10,3508)(20,4225)
    };
\addplot[
    color=blue,
    mark=triangle,
    ]
    coordinates {
    (1,570)(2,511)(5,463)(10,545)(20,501)
    };
\addplot[
    color=olive,
    mark=diamond,
    ]
    coordinates {
    (1,1383)(2,1389)(5,1376)(10,1369)(20,1371)
    };  
\addplot[
    color=orange,
    mark=pentagon,
    ]
    coordinates {
    (1,401)(2,446)(5,480)(10,501)(20,620)
    };
    \tiny
    %{{\legend{InfluxDB,OpenTSDB,TimescaleDB,KairosDB}}}
    {{\legend{I,O,T,K}}}
\end{axis}
\end{tikzpicture}

%% file: draws/ingestion/space.tex
\begin{tikzpicture}
\pgfplotsset{width=5.55cm, compat=1.9}
\begin{axis}[
% 	x tick label style={/pgf/number format/1000 sep=},
% 	ylabel={Data used disk space [byte]},
% 	xlabel={Time series data distribution type},
% 	enlargelimits=0.05,
% 	legend pos=north west,
% 	symbolic x coords={Inf,Ope,Tim,Kai},
%     xtick=data,
% 	ybar interval=0.7,
	x tick label style={/pgf/number format/1000 sep=},
	ylabel={Data used disk space [byte]},
	xlabel={Time series database},
	enlargelimits=0.15,
	legend pos=north west,
	symbolic x coords={I,T,K,O},
    xtick=data,
	%ybar interval=0.7,
 	ybar=2pt,% configures ‘bar shift’
    bar width=1pt,
    ymajorgrids=true,
    grid style=dashed,
    ybar
    % x tick label style={anchor=east},
]
\addplot
	coordinates {
	(I, 20556)  (T, 1599580)
	(K, 1662100) (O, 6170376)
	};
\addplot
	coordinates {
	(I, 28360)  (T, 1599552)
	(K, 1664900) (O, 10118928)
	};
\addplot
	coordinates {
	(I, 544428)  (T, 1550984)
	(K, 1796360) (O, 10369580)
	};
\addplot
	coordinates {
	(I, 671380)  (T, 1599648)
	(K, 2465464) (O, 10046568)
	};
\addplot
	coordinates {
	(I, 703080)  (T, 1599104)
	(K, 2744264) (O, 10382388)
	};
\tiny{{\legend{All Constant,All Square,All Sawtooth,All Sine,All Random}}}
\end{axis}
\end{tikzpicture}

%%%%%%%%%%%%%%%%%%%%%%%%%%%方案二%%%%%%%%%%%%%%%%%%%%%%%%%%%%%%%%%%%
% \begin{tikzpicture}
% \begin{axis}[
% 	x tick label style={/pgf/number format/1000 sep=},
% 	ylabel={Throughput [thousand points per second]},
% 	xlabel={Time series data distribution type},
% 	enlargelimits=0.1,
% 	legend pos=north west,
% 	symbolic x coords={constant,square,sawtooth,sine,radom},
%     xtick=data,
% 	%ybar interval=0.7,
%  	ybar=2pt,% configures ‘bar shift’
%     bar width=3pt,
%     ybar
%     % x tick label style={anchor=east},
% ]
% \addplot
% 	coordinates {
% 	(constant, 20556) (square, 28360)
% 	(sawtooth, 544428)(sine, 671380) (radom, 703080)
% 	};
% \addplot
% 	coordinates {
% 	(constant, 6170376) (square, 10118928)
% 	(sawtooth, 10369580)(sine, 10046568) (radom, 10382388)
% 	};
% \addplot
% 	coordinates {
% 	(constant, 1599580) (square, 1599552)
% 	(sawtooth, 1550984)(sine, 1599648) (radom, 1599104)
% 	};
% \addplot
% 	coordinates {
% 	(constant, 1662100) (square, 1664900)
% 	(sawtooth, 1796360)(sine, 2465464) (radom, 2744264)
% 	};
% \tiny{{\legend{InfluxDB,OpenTSDB,TimescaleDB,KairosDB}}}
% \end{axis}
% \end{tikzpicture}

%% file: draws/ingestion/step.tex
\begin{tikzpicture}
\pgfplotsset{width=5.55cm, compat=1.9}
\begin{axis}[
% 	x tick label style={/pgf/number format/1000 sep=},
% 	ylabel={Data used disk space [byte]},
% 	xlabel={Time series data distribution type},
% 	enlargelimits=0.05,
% 	legend pos=north west,
% 	symbolic x coords={Inf,Ope,Tim,Kai},
%     xtick=data,
% 	ybar interval=0.7,
	x tick label style={/pgf/number format/1000 sep=},
	ylabel={Data used disk space [byte]},
	xlabel={Time series database},
	enlargelimits=0.15,
	legend pos=north west,
	symbolic x coords={I,T,K,O},
    xtick=data,
	%ybar interval=0.7,
 	ybar=2pt,% configures ‘bar shift’
    bar width=4pt,
    ymajorgrids=true,
    grid style=dashed,
    ybar
    % x tick label style={anchor=east},
]
\addplot
	coordinates {
	(I, 532116)  (T, 1602404)
	(K, 1725000) (O, 13997284)
	};
\addplot
	coordinates {
	(I, 326160)  (T, 1600008)
	(K, 1625052) (O, 10534592)
	};
\tiny{{\legend{Random Interval,Uniform Interval}}}
\end{axis}
\end{tikzpicture}

% \begin{tikzpicture}
% \begin{axis}[
% % 	x tick label style={/pgf/number format/1000 sep=},
% % 	ylabel={Data used disk space [byte]},
% % 	xlabel={Time series data distribution type},
% % 	enlargelimits=0.05,
% % 	legend pos=north west,
% % 	symbolic x coords={Inf,Ope,Tim,Kai},
% %     xtick=data,
% % 	ybar interval=0.7,
% 	x tick label style={/pgf/number format/1000 sep=},
% 	ylabel={Throughput [points per second]},
% 	xlabel={Time series database},
% 	enlargelimits=0.15,
% 	legend pos=north east,
% 	symbolic x coords={Inf,Tim,Kai,Ope},
%     xtick=data,
% 	%ybar interval=0.7,
%  	ybar=2pt,% configures ‘bar shift’
%     bar width=4pt,
%     ymajorgrids=true,
%     grid style=dashed,
%     ybar
%     % x tick label style={anchor=east},
% ]
% \addplot
% 	coordinates {
% 	(Inf, 1955684)  (Tim, 672321)
% 	(Kai, 575066) (Ope, 936994)
% 	};
% \addplot
% 	coordinates {
% 	(Inf, 1978381)  (Tim, 673990)
% 	(Kai, 579703) (Ope, 946994)
% 	};
% \tiny{{\legend{Random Interval,Uniform Interval}}}
% \end{axis}
% \end{tikzpicture}

%% file: draws/ingestion/overflow.tex
\begin{tikzpicture}
\pgfplotsset{
    every axis legend/.append style={
    at={(0.5,1.05)},
    anchor=south
    }, 
}
\begin{axis}[
    legend columns=4,
    xlabel={Ratio of out-of-order data [\%]},
    ylabel={Throughput [thousand points per second]},
    xmin=10, xmax=50,
    ymin=0, ymax=2000,
    xtick={10,20,30,40,50},
    ytick={0,500,1000,1500,2000},
    % legend pos=south west,
    ymajorgrids=true,
    grid style=dashed,
]
\addplot[
    color=purple,
    mark=square,
    ]
    coordinates {
    (10,1036)(20,1035)(30,1034)(40,1034)(50,1033)
    };
\addplot[
    color=blue,
    mark=triangle,
    ]
    coordinates {
    (10,502)
    };
\addplot[
    color=olive,
    mark=diamond,
    ]
    coordinates {
    (10,1409)(20,1375)(30,1412)(40,1350)(50,1398)
    };  
\addplot[
    color=orange,
    mark=pentagon,
    ]
    coordinates {
    (10,563)(20,528)(30,524)(40,428)(50,426)
    };
    % \tiny{{\legend{InfluxDB,OpenTSDB,TimescaleDB,KairosDB}}}
    \tiny{{\legend{I,O,T,K}}}
\end{axis}
\end{tikzpicture}

%% file: draws/precise_pic.tex
\begin{tikzpicture}
\pgfplotsset{
    every axis legend/.append style={
    at={(0.5,1.05)},
    anchor=south
    }, 
}
\begin{axis}[
    legend columns=4,
    xlabel={Number of series per query},
    ylabel={Cost-time [ms per query]},
    xmin=20, xmax=100,
    ymin=0, ymax=100,
    xtick={0,20,40,60,80,100},
    ytick={0,20,40,60,80,100},
    %legend pos=north west,
    ymajorgrids=true,
    grid style=dashed,
]
\addplot[
    color=purple,
    mark=square,
    ]
    coordinates {
    (20,3.3)(40,4.4)(60,4.8)(80,5.6)(100,6.2)
    };
\addplot[
    color=blue,
    mark=triangle,
    ]
    coordinates {
    (20,5.9)(40,7.4)(60,9.1)(80,11.4)(100,12.7)
    };
\addplot[
    color=olive,
    mark=diamond,
    ]
    coordinates {
    (20,1.57)(40,1.67)(60,1.79)(80,1.89)(100,1.98)
    };  
\addplot[
    color=orange,
    mark=pentagon,
    ]
    coordinates {
    (20,75)(40,75)(60,76)(80,76)(100,77)
    };
    \tiny
    %{{\legend{InfluxDB,OpenTSDB,TimescaleDB,KairosDB}}}
    {{\legend{I,O,T,K}}}
\end{axis}
\end{tikzpicture}

%% file: draws/range_pic.tex
\begin{tikzpicture}
\pgfplotsset{
    every axis legend/.append style={
    at={(0.5,1.05)},
    anchor=south
    }, 
}
\begin{axis}[
    legend columns=4,
    xlabel={Result point number per query},
    ylabel={Cost-time [ms per query]},
    xmin=500, xmax=2500,
    ymin=0, ymax=600,
    xtick={500,1000,1500,2000,2500},
    ytick={0,100,200,300,400,500,600},
    %legend pos=north west,
    ymajorgrids=true,
    grid style=dashed,
]
\addplot[
    color=purple,
    mark=square,
    ]
    coordinates {
    (500,5.7)(1000,9.7)(1500,13.1)(2000,18.0)(2500,21.9)
    };
\addplot[
    color=blue,
    mark=triangle,
    ]
    coordinates {
    (500,108)(1000,320)(1500,370)(2000,413)(2500,500)
    };
\addplot[
    color=olive,
    mark=diamond,
    ]
    coordinates {
    (500,88)(1000,168)(1500,213)(2000,280)(2500,347)
    };  
\addplot[
    color=orange,
    mark=pentagon,
    ]
    coordinates {
    (500,10)(1000,13)(1500,16)(2000,19)(2500,23)
    };
    \tiny
    %{{\legend{InfluxDB,OpenTSDB,TimescaleDB,KairosDB}}}
    {{\legend{I,O,T,K}}}
\end{axis}
\end{tikzpicture}

%% file: draws/collumn_pic.tex
\begin{tikzpicture}
\pgfplotsset{
    every axis legend/.append style={
    at={(0.5,1.05)},
    anchor=south
    }, 
}
\begin{axis}[
    legend columns=4,
    xlabel={Series number per query},
    ylabel={Cost-time [ms per query]},
    xmin=10, xmax=50,
    ymin=0, ymax=1600,
    xtick={10,20,30,40,50},
    ytick={0,200,400,600,800,1000,1200,1400,1600},
    % legend pos=north west,
    ymajorgrids=true,
    grid style=dashed,
]
\addplot[
    color=purple,
    mark=square,
    ]
    coordinates {
    (10,276)(20,556)(30,836)(40,1137)(50,1401)
    };
\addplot[
    color=blue,
    mark=triangle,
    ]
    coordinates {
    (10,501)(20,696)(30,822)(40,938)(50,1038)
    };
\addplot[
    color=olive,
    mark=diamond,
    ]
    coordinates {
    (10,364)(20,421)(30,475)(40,548)(50,598)
    };  
\addplot[
    color=orange,
    mark=pentagon,
    ]
    coordinates {
    (10,739)(20,865)(30,1059)(40,1203)(50,1406)
    };
    \tiny
    %{{\legend{InfluxDB,OpenTSDB,TimescaleDB,KairosDB}}}
    {{\legend{I,O,T,K}}}
\end{axis}
\end{tikzpicture}

%% file: draws/count_agg_pic.tex
\begin{tikzpicture}
\pgfplotsset{
    every axis legend/.append style={
    at={(0.5,1.05)},
    anchor=south
    }, 
}
\begin{axis}[
    legend columns=4,
    xlabel={Number of series per query},
    ylabel={Cost-time [ms per query]},
    xmin=20, xmax=100,
    ymin=0, ymax=1600,
    xtick={0,20,40,60,80,100},
    ytick={0,200,400,600,800,1000,1200,1400,1600},
    %legend pos=north west,
    ymajorgrids=true,
    grid style=dashed,
]
\addplot[
    color=purple,
    mark=square,
    ]
    coordinates {
    (20,31.0)(40,60.2)(60,87.8)(80,115.8)(100,143.5)
    };
\addplot[
    color=blue,
    mark=triangle,
    ]
    coordinates {
    (20,382)(40,445)(60,501)(80,527)(100,544)
    };
\addplot[
    color=olive,
    mark=diamond,
    ]
    coordinates {
    (20,394)(40,495)(60,578)(80,665)(100,750)
    };  
\addplot[
    color=orange,
    mark=pentagon,
    ]
    coordinates {
    (20,703)(40,815)(60,930)(80,1230)(100,1370)
    };
    \tiny
    %{{\legend{InfluxDB,OpenTSDB,TimescaleDB,KairosDB}}}
    {{\legend{I,O,T,K}}}
\end{axis}
\end{tikzpicture}

%% file: draws/max_agg_pic.tex
\begin{tikzpicture}
\pgfplotsset{
    every axis legend/.append style={
    at={(0.5,1.05)},
    anchor=south
    }, 
}
\begin{axis}[
    legend columns=4,
    xlabel={Number of series per query},
    ylabel={Cost-time [ms per query]},
    xmin=20, xmax=100,
    ymin=0, ymax=1600,
    xtick={0,20,40,60,80,100},
    ytick={0,200,400,600,800,1000,1200,1400,1600},
    %legend pos=north west,
    ymajorgrids=true,
    grid style=dashed,
]
\addplot[
    color=purple,
    mark=square,
    ]
    coordinates {
    (20,31.9)(40,62.2)(60,92.6)(80,122.8)(100,158.7)
    };
\addplot[
    color=blue,
    mark=triangle,
    ]
    coordinates {
    (20,332)(40,395)(60,500)(80,528)(100,548)
    };
\addplot[
    color=olive,
    mark=diamond,
    ]
    coordinates {
    (20,395)(40,496)(60,579)(80,666)(100,750)
    };  
\addplot[
    color=orange,
    mark=pentagon,
    ]
    coordinates {
    (20,706)(40,832)(60,944)(80,1370)(100,1500)
    };
    \tiny
    %{{\legend{InfluxDB,OpenTSDB,TimescaleDB,KairosDB}}}
    {{\legend{I,O,T,K}}}
\end{axis}
\end{tikzpicture}

%% file: draws/groupby_pic.tex
\begin{tikzpicture}
\pgfplotsset{
    every axis legend/.append style={
    at={(0.5,1.05)},
    anchor=south
    }, 
}
\begin{axis}[
    legend columns=4,
    xlabel={Group by time range span [\%]},
    ylabel={Cost-time [ms per query]},
    xmin=0.1, xmax=10,
    ymin=0, ymax=1800,
    xtick={0.1,1,2,5,10},
    ytick={0,300,600,900,1200,1500,1800},
    %legend pos=north east,
    ymajorgrids=true,
    grid style=dashed,
]
\addplot[
    color=purple,
    mark=square,
    ]
    coordinates {
    (0.1,97)(1,23)(2,18)(5,16)(10,15)
    };
\addplot[
    color=blue,
    mark=triangle,
    ]
    coordinates {
    (0.1,440)(1,430)(2,429)(5,427)(10,428)
    };
\addplot[
    color=olive,
    mark=diamond,
    ]
    coordinates {
    (0.1,381)(1,381)(2,381)(5,380)(10,380)
    };  
\addplot[
    color=orange,
    mark=pentagon,
    ]
    coordinates {
    (0.1,1766)(1,793)(2,736)(5,703)(10,699)
    };
    \tiny%{{\legend{InfluxDB,OpenTSDB,TimescaleDB,KairosDB}}}
    {{\legend{I,O,T,K}}}
\end{axis}
\end{tikzpicture}